\documentclass[aps,prl,twocolumn,showpacs,amsmath,amssymb,superscriptaddress,floatfix]{revtex4}
\usepackage{graphicx,epsf}
\usepackage{bm}

\begin{document}

\title{Valley polarization effects on the localization in graphene Landau levels}

\author{A. L. C. Pereira}

\author{P. A. Schulz}
\address{Instituto de F\'\i sica Gleb Wataghin, Unicamp,  C.P. 6165, 13083-970, Campinas,
Brazil}

\date{\today}

\begin{abstract}
 Effects of disorder and valley polarization in graphene are investigated in
the quantum Hall regime. We find anomalous localization properties for the lowest Landau level (LL),
where disorder can induce wavefunction delocalization
(instead of localization), both for white-noise and gaussian-correlated disorder.
We quantitatively identify the contribution of each sublattice to wavefunction amplitudes.
Following the valley (sublattice) polarization of states within LLs for increasing disorder we show:
(i) valley mixing in the lowest LL is the main
effect behind the observed anomalous localization properties, (ii) the polarization suppression with increasing disorder depends on the localization for the white-noise model, while, (iii) the disorder induces a
partial polarization in the higher Landau levels for both disorder models.
\end{abstract}


% 73.43.-f: QHE
% 73.23.-b: Electronic transport in mesoscopic systems
% 73.63.-b: Electronic transport in nanoscale materials and structures

\maketitle

%%%%% Introduction %%%%%%%%%%%%%%%%%%%%%%%

Graphene, a condensed matter stage for quantum electrodynamics, has
attracted an increasing amount of interest and now monolayer graphite
has turned into the most versatile electronic system under study \cite{geim}.
In the presence of magnetic field $B$, an anomalous integer quantum Hall
effect (QHE) \cite{novoselov,zhang1} is observed in graphene, where
the sequence of plateaus is shifted by half-integer if compared
to the usual QHE, as have been theoretically predicted
\cite{previsao1,previsao2}. The anomalous sequence of plateaus is
given by: $\sigma_{xy}=(4e^2/h)(n+1/2)$, where integer $n$ is the Landau
level (LL) index and the factor 4 comes from spin and valley degeneracies.

The unique feature of a Dirac-cone-like band structure in the energy
range of interest (Fermi energy) has led to an avalanche of
intriguing observations and a bunch of novel predictions
\cite{geim}. There are, in particular, two important discussions
taking place recently in the literature on graphene: (i) First, the
discussion about the role played by valley degree of freedom -
started after the observation \cite{zhang2} of the lifting of valley
degeneracy for the lowest ($n$=0) LL at high magnetic fields. Many theoretical works have been
studying the mechanisms related
to valley polarization
\cite{alicea,lederer,levitov,ando,altshuler,sheng2,chak} and
proposing possible applications \cite{beenakker1,niu,beenakker2}.
(ii) Second, the discussion about anomalous phenomena on the
conduction and localization around Dirac point. On the experimental
side, the suppression of weak localization was observed in graphene
\cite{morozov}, while on the theoretical side, intriguing
observations of disorder causing the increase (rather than decrease)
of the transmission at the Dirac point have been
reported\cite{mirlin,beenakker,nomura,titov}, still without a
consensus about the origins of this striking mechanism.

We report here findings in both (i) and (ii) subjects of discussion mentioned
above and also a connection between them, as we found that mixing of
valley polarized states can explain the anomalous localization we
observe for graphene systems in the quantum Hall regime. Related to this context,
although not for graphene systems, we call attention to a recent experimental
report on the influence of valley polarization on the transport properties of a two-dimensional electron
system \cite{gunawan}.

%%%%% Model %%%%%%%%%%%%%%%%%%%%%%%

The fascinating graphene landscape can be widely explored within a
single electron framework.
The energy dispersion relation of graphene is linear
around the two inequivalent Brillouin zone corners $K$ and $K'$, forming
two cones in the band structure, whose apex is named Dirac-point.
We calculate here the density of states (DOS),
localization properties and the sublattice polarization of wave functions around Dirac point
in the presence of a perpendicular magnetic field,
within a tight-binding model description.
The graphene band structure is already well described by a nearest neighbor hopping between s-like orbitals
in a honeycomb lattice: $H = \sum_{i} \varepsilon_{i} c_{i}^{\dagger} c_{i}
+ t  \sum_{<i,j>} (e^{i\phi_{ij}} c_{i}^{\dagger} c_{j} + e^{-i\phi_{ij}}
c_{j}^{\dagger} c_{i})$, where $c_{i}$ is the fermionic operator on site $i$.
The magnetic field is introduced by means of the phase
$\phi_{ij}= 2\pi(e/h) \int_{j}^{i} \mathbf{A} \! \cdot \! d \mathbf{l} \;$ in
the hopping parameter $t$ ($t$$\approx$2.7eV for graphene, but we give
results parameterized by $t$). In the Landau gauge, $\phi_{ij}\!=\!0$
along the $x$ direction and $\phi_{ij}\!=\pm \pi (x/a) \Phi / \Phi_{0}$
along the $\mp y$ direction, with $\Phi / \Phi_{0}=Ba^{2}\sqrt{3}e/h$
 ($a$=2.46{\AA} is the lattice constant for graphene).
Here we address two on-site disorder models: white-noise fluctuations -
emulating short-range scatterers - and correlated (smoothed)
disorder landscapes \cite{pereira}. For the white-noise model,
uncorrelated orbital energies are sorted within
$\varepsilon_{i} \leq |W/2|$, while for the correlated
disorder, a gaussian correlation $\varepsilon_{i} =
\frac{1}{\pi \lambda^{2}} \sum_{j} \varepsilon_{j}
e^{-|\mathbf{R}_{i}-\mathbf{R}_{j}|^{2}/\lambda^2}$ is assumed,
with correlation length $\lambda$ and  $\varepsilon_{j} \leq |W/2|$.
The importance of the disorder in these systems is widely accepted and addressed
by different groups \cite{previsao2,macdonald,mirlin2,castroneto1,castroneto3,brey,sheng}.

The degree of localization of the states is estimated by the
participation ratio, $PR=1/(N \sum_{i=1}^{N}|\psi_{i}|^{4})$,
where $\psi_{i}$ is the amplitude of the normalized wavefunction on site $i$, and $N$
is the total number of lattice sites \cite{thouless}.
All the results shown here have been calculated on disordered unit
cells of 60$\times$60 sites (hexagonal lattice of
60 zig-zag chains, each containing 60 sites),
with periodic boundary conditions. Averages over hundred disorder realizations
are undertaken. For the correlated disorder we take $\lambda$=$3a$.
We consider a low magnetic flux through the lattice unit cell,
$\Phi/\Phi_0 \!=\! 1/30$,  and only few LLs, to keep within the continuum
limit of the Hofstadter spectrum for the hexagonal lattice \cite{aoki} .

%%%%% Figure 1 %%%%%%%%%%%%%%%%%%%%%%%

\begin{figure}[t]
%\vspace{-0.5cm}
\centerline{\includegraphics[width=8.5cm]{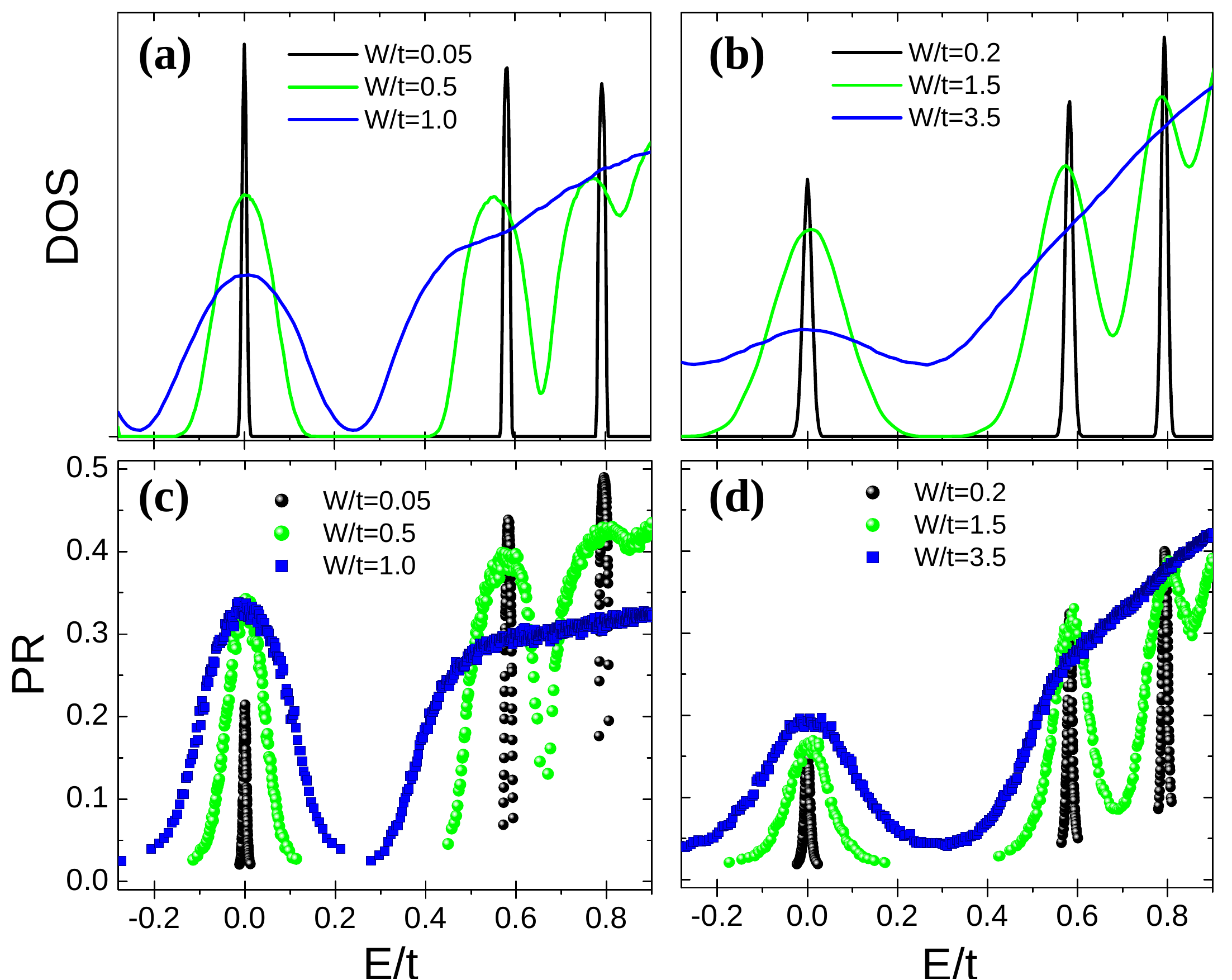}}
%\vspace{-0.5cm}
\caption{(color online) (a)(b) DOS
for different disorder strengths (broadening of LLs increases with
disorder) and (c)(d) the corresponding participation ratio of
states. First column (a)(c) corresponds to the white-noise
disorder and the second (b)(d) to correlated disorder
with $\lambda$=3$a$. Three LLs are shown ($n$=0,1,2)
for flux $\Phi / \Phi_0$=$1/30$. Localization
properties (observed from the evolution of PR with disorder) are
very different in the $n$=0 LL compared to the others.} \label{Fig1}
\end{figure}

%%%%% Discussing Fig. 1 %%%%%%%%%%%%%%%%%%%%%%%

Figure 1 shows the DOS and participation ratio (PR)
around the Dirac point for increasing disorder in two different models: white-noise (Fig.1a and 1c), and
correlated disorder (Fig.1b and 1d). From the DOS,
one can see that the energy range shown corresponds, for this magnetic flux, to the
lowest three LLs ($n$=0,1,2), well resolved  at small disorder amplitudes.
The spectrum is symmetrical around zero energy, so we are not showing the hole side ($n$=-1,-2).
The PR, by definition \cite{thouless}, gives the proportion of the $N$ lattice sites over which
the wavefunction is spread. In this way, the smaller
the value of PR, the more localized the state is. In fact, a characteristic localization is observed
at LL tails, while extended states are identified
by peaks in the PR around the center of each LL \cite{pereira}. We first want to call attention in Fig.1 to the effect of
levitation of extended states \cite{laughlin,sheng}. Increasing disorder leads
to LL broadening (as expected due to intra
LL mixing) as well as to a Landau band repulsion, due to inter
LL mixing \cite{uemura,aldea}. This band repulsion
can be clearly seen for the DOS of the LL $n$=1, as a shift of the Landau band in direction
to E=0. Looking at the PRs, one can observe that the delocalized states
of the $n$=1 LL do not move towards lower energies as fast as the DOS, characterizing an effective levitation of extended states,
as for Schroedinger-like carriers at the electronic structure extremes \cite{pereira}. Such levitation is necessarily frustrated at the $n$=0 LL.

The main effect we want to point out in Fig.1 is an anomalous increase of PR with increasing disorder, occurring for states of the $n$=0 LL. This can be observed in both disorder models, but is more evident for the white-noise case. From Fig.1c, one easily observes PR peaks corresponding to the $n$=1 and $n$=2 LLs having smaller values with increasing disorder, as expected, while for the $n$=0 LL, PR around the peak is substantially increased for disorder amplitudes from $W/t$=0.05 to 0.5 (PR values at $n$=0 LL peak are followed as a function of disorder in Figures 5 and 6, so more details are discussed further in the text). The observed increase in the PR is surprising and called anomalous because increasing disorder is expected to have the opposite effect: making states more localized. The same anomalous effect persists for larger system sizes (not shown here). Nevertheless it is worth mentioning that size effects would not be able to explain this observed increase of PR wi
 t
 h increasing disorder. In this way, one has to look at another degree of freedom in the problem.

%%%%% Levitation %%%%%%%%%%%%%%%%%%%%%%%

%%%%%%%%%%%%%%%%%%%%%%%%%%%%%%%%%%%%%%%%

%%%%% Figure 2 %%%%%%%%%%%%%%%%%%%%%%%

\begin{figure}[b]
%\vspace{-1.0cm}
\centerline{\includegraphics[width=8.4cm,height=5cm]{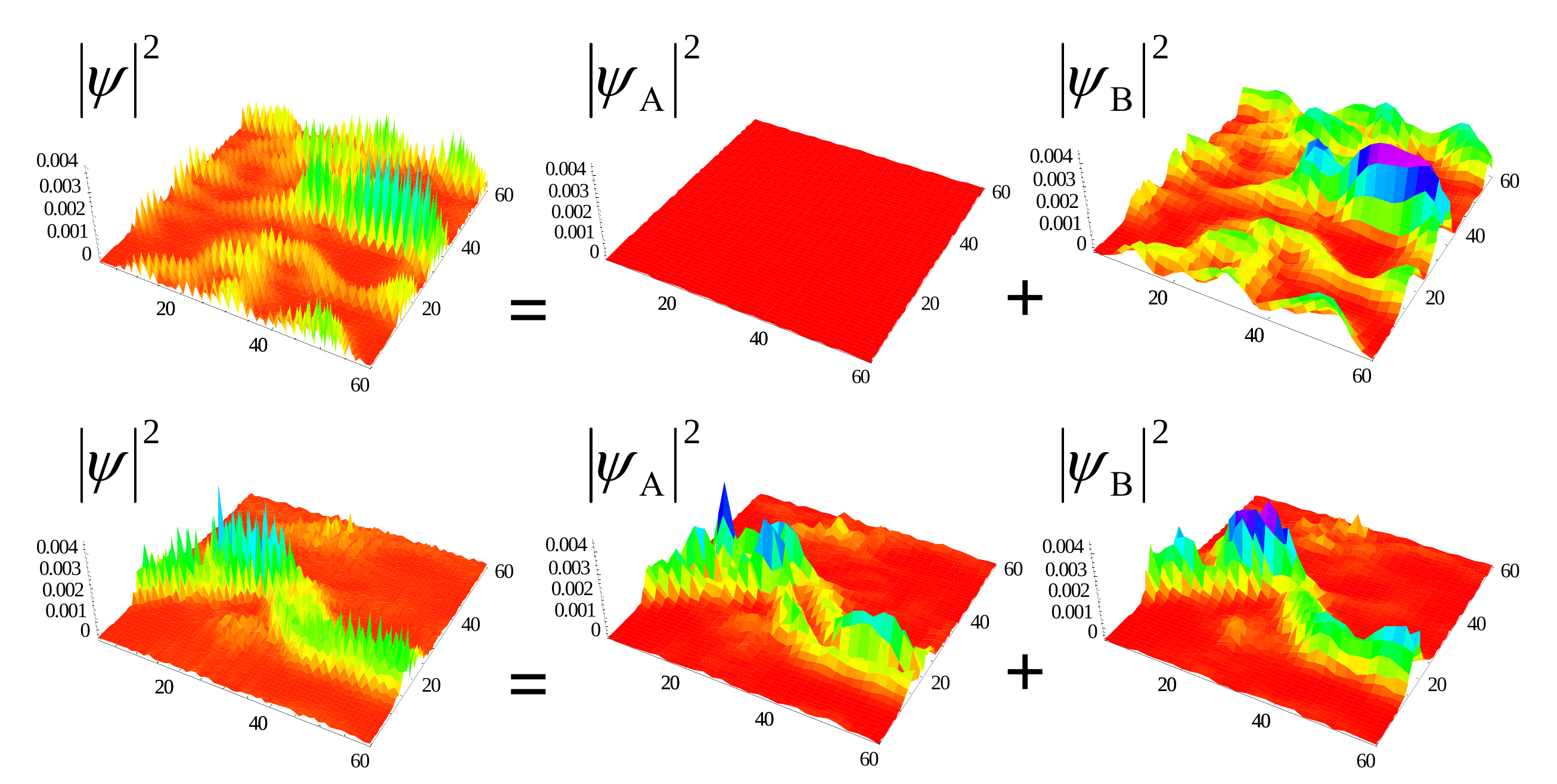}}
%\vspace{-1.3cm}
\caption{ (color online)
Wavefunction, and its sublattice decomposition, of a valley
polarized state from $n$=0 LL (top) and a valley unpolarized state
from $n$=1 LL (bottom). Wavefunction amplitudes  ($|\Psi|^2$) are firstly shown
over all the 60$\times$60 lattice sites, and then contributions from each
sublattice ($|\Psi_A|^2$ and $|\Psi_B|^2$) are plotted separately.
Results are for a correlated disorder with
$W/t$=$1.0$, $\lambda$=3$a$ and $\Phi / \Phi_0$=$1/30$.} \label{Fig2}
\end{figure}

%%%%% Discussing Fig. 2 %%%%%%%%%%%%%%%%%%%%%%%

A unit cell of the graphene hexagonal lattice
contains two carbon atoms, defining two sublattices, $A$ and $B$.
For the ideal graphene system (no disorder) in the presence of magnetic field,
there is a great
difference between the wavefunctions from the $n$=0 LL  and that from higher levels:
in the lowest LL, wavefunctions have non-zero amplitudes on only one of the sublattices \cite{ando}.
But how does disorder affect this picture? How does the sublattice polarization of the $n$=0 LL evolves as the disorder increasingly introduces Landau level coupling? Do the higher LLs remain unpolarized with increasing disorder? 
 We show next that the role of disorder in destroying sublattice polarization in the $n$=0 LL reveals to be related to the anomalous behavior of the PR at $n$=0 LL.

We start showing in Figure 2 examples of wavefunctions probability densities ($|\Psi|^2$) and their decompositions in the two sublattices, $|\Psi_A|^2$ and $|\Psi_B|^2$. We define here $|\Psi_{A(B)}|^2$ as the sum of wavefunction amplitudes only over the sites of sublattice A(B). In this way, due to
wavefunction normalization, $|\Psi_A|^2 +|\Psi_B|^2 = 1$. The two decomposed states shown in Fig.2 correspond to typical extended states, one from the
$n$=0 LL (top) and the other from the $n$=1 LL (bottom), for a correlated disorder potential.
We can clearly see that the state from $n$=0 LL is sublattice (or valley)
polarized for the small disorder amplitude considered here: the sublattice A contribution to the chosen state is still very small and can be neglected (all plots have the same amplitude scale). On the other hand, the state from the $n$=1 LL is
valley unpolarized: by eye inspecting the probability densities,
the contribution of both sublattices are equally relevant.

%%%%% Discussing Fig. 3 - Correlated %%%%%%%%%%%%%%%%%%%%%%%

The distribution of wavefunction amplitudes over only one of the graphene sublattices, or preferentially over one of them, is what we call here sublattice (valley) polarization.
To analyze the limits where this polarization takes place and address the questions posed above, we define a simple quantity
to infer the valley polarization:

\begin{equation}
Polarization = ||\Psi_{A}|^2-|\Psi_{B}|^2|
\end{equation}

%\begin{equation}
%|\Psi_{A(B)}|^2 \equiv  \negthickspace \negthickspace
%\sum_{\substack{i=1 \\ i \: \in  \: \text{sublat.A(B)}}}^{N} \negthickspace
%\negthickspace \negthickspace \negthickspace \negthickspace
%|\psi_{i}|^{2} {\Big /} \sum_{i=1}^{N}|\psi_{i}|^{2}
%\end{equation}

%%%%% Figure 3 %%%%%%%%%%%%%%%%%%%%%%%

\begin{figure}[b]
%\vspace{-0.5cm}
\centerline{\includegraphics[width=8.4cm]{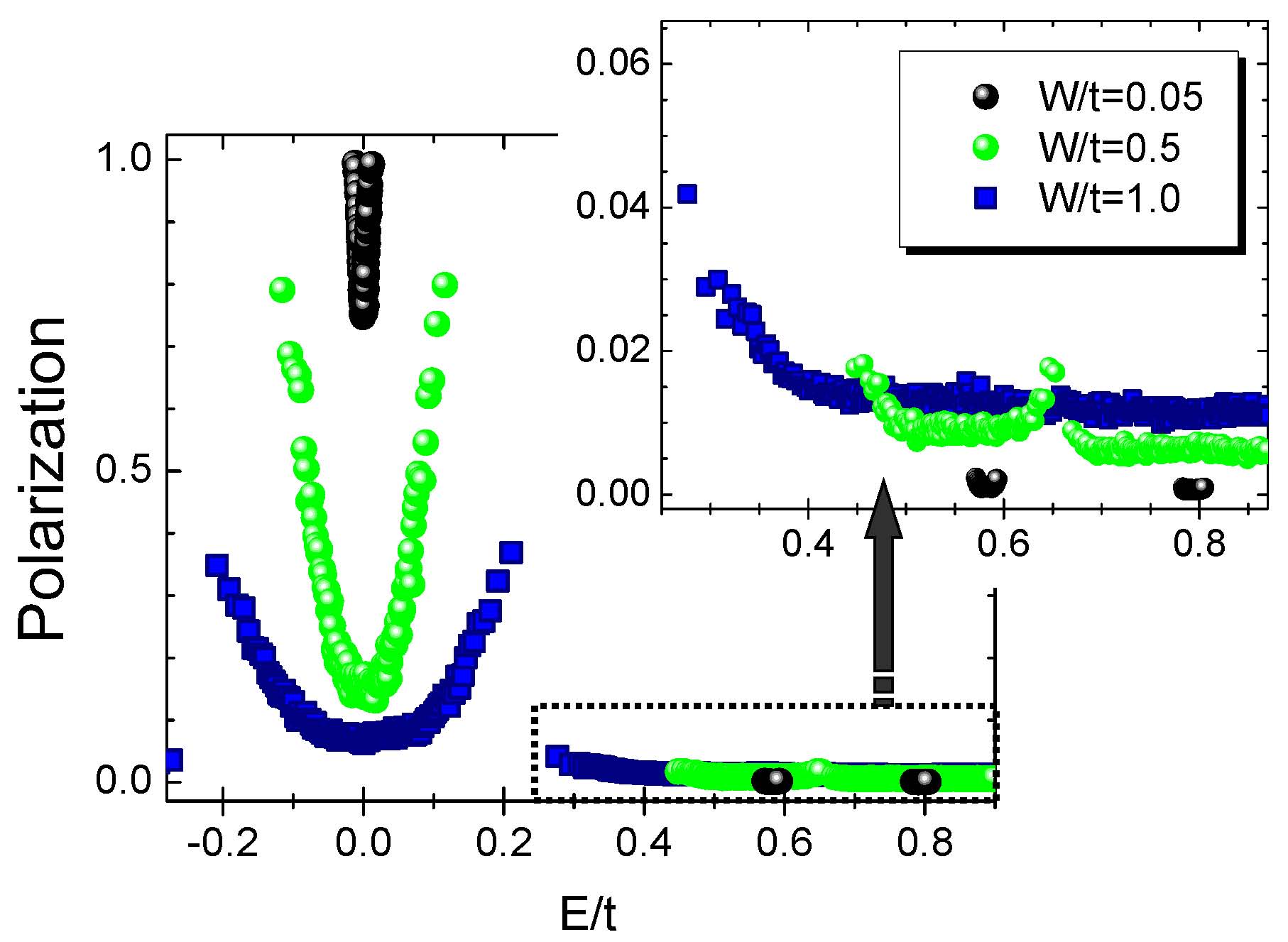}}
%\vspace{-0.6cm}
\caption{(color online) Valley polarization of the states within $n$=0,1,2 LLs (same energy range shown in Fig.1), for three different disorder amplitudes in the \underline{white-noise disorder model}. In the detail, the zoom of the region inside the dashed line, corresponding to the fine structure of the polarization in the $n$=1 and $n$=2 LLs.}
\label{Fig3}
\end{figure}

\hspace{-\parindent}This quantity gives the difference in modulus of the contribution from each sublattice to the total probability density of the state. Zero polarization means that the state is equally distributed on both sublattices, while total polarization, corresponding to values 1.0, means that the state is completely on sublattice A or B.

With this definition, valley polarization of the states can be followed for increasing
disorder in the white-noise and correlated models - the results are shown in Figures 3 and 4, respectively. Note that the main graphs have  exactly the same energy range shown in Fig.1. Zoom graphs show the surprising fine structure of polarization on the $n$=1 and $n$=2 LLs.
Comparing disorder models for equivalent LL broadening, it is clear that white-noise disorder (Fig.3) induces more valley mixing on the $n$=0 LL than the correlated disorder (Fig.4). This effect is in connection to the fact that the larger the length scale of the disorder compared to the lattice constant, the smaller the capacity of disorder to cause inter-valley mixing \cite{ando}. However, the wavefunctions polarization observed from Figs. 3 and 4 show other intriguing fingerprints that go beyond natural expectations, as we discuss next.

%%%%% Figure 4 %%%%%%%%%%%%%%%%%%%%%%%

\begin{figure}[b]
%\vspace{-0.5cm}
\centerline{\includegraphics[width=8.4cm]{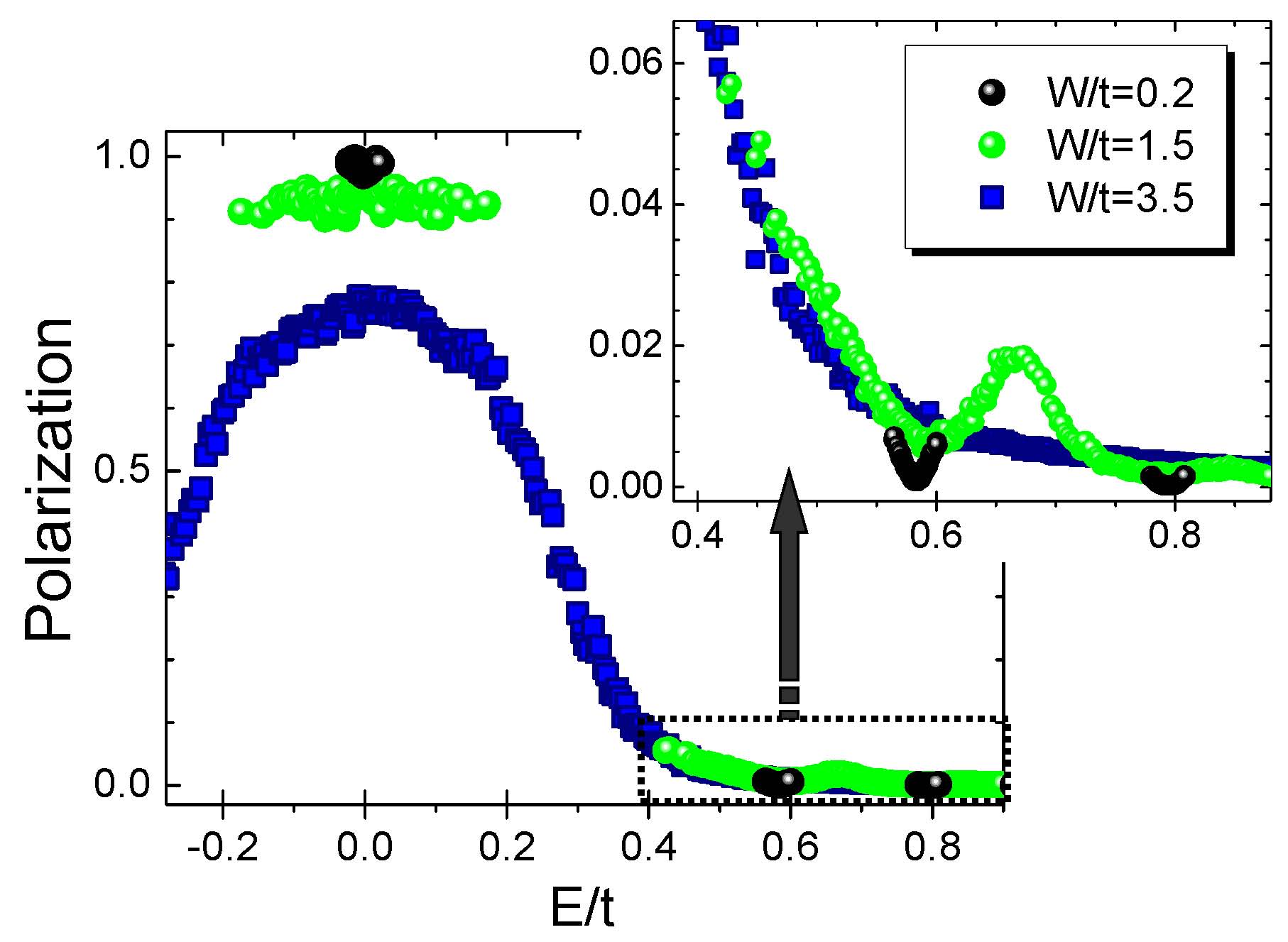}}
%\vspace{-0.8cm}
\caption{(color online) Valley polarization of the states within $n$=0,1,2 LLs (same energy range shown in Fig.1), for three different disorder amplitudes in the \underline{gaussian-correlated disorder model} ($\lambda$=3$a$). In the detail, the zoom of the region inside the dashed line, corresponding to the fine structure of the polarization in the $n$=1 and $n$=2 LLs.}
\label{Fig4}
\end{figure}

For white-noise model (Fig.3), the states from the $n$=0 band clearly show an interesting ``{\it U-shape}" polarization profile: states at the band tails - strongly localized (as observed in Fig.1c) - remain more polarized than the delocalized ones, from LL center, which become increasingly unpolarized with disorder. This relation between localization length and polarization of the states in the $n$=0 LL is not present when the disorder potential is smoother (Fig.4). From Fig.4 we see that the destruction of the sublattice polarization with Gaussian-correlated disorder is a slow process, that affects all states from the $n$=0 LL in the same manner, regardless of being localized or delocalized.

Another striking observation that emerges from this analysis is the induced polarization beyond the $n$=0 LL: states that are completely unpolarized in the absence of disorder, start to show an
increasing partial polarization (up to $\approx$6$\%$) in the disordered limit. Zoomed graphs show that the fine structure of this $n>0$  polarization presents the ``{\it U-shape}" polarization profiles for both disorder models, observed for both $n$=1 and $n$=2 LLs.

The results shown in Figs. 3 and 4 are averages over hundreds of disorder configurations. However, focusing on individual disorder realizations one observes that (not shown here) the polarization may jump randomly from one sublattice to the other for neighbour states in energy within the LL. This behaviour is inferred from the quantity described in eq.(1) when the modulus is skiped. 

%%%%% Figure 5 %%%%%%%%%%%%%%%%%%%%%%%

\begin{figure}[b]
\vspace{-0.2cm}
\includegraphics[width=9.3cm]{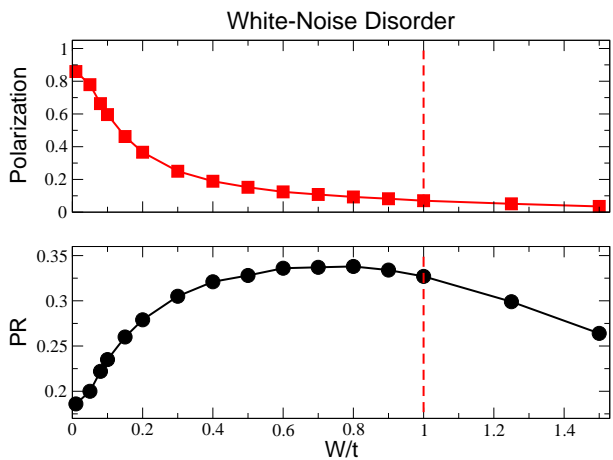}
\vspace{-0.2cm}
\caption{(color online) Values of valley polarization and participation ratio (PR) for the states at Dirac point (at the center of the $n$=0 LL), as a function of white-noise disorder amplitude. Dashed line indicates the $W/t$ for which the broadening of LLs is such that the $n$=0 LL starts to overlap with neighbors bands.}
\label{Fig4}
\end{figure}

%%%%% Figure 6 %%%%%%%%%%%%%%%%%%%%%%%

\begin{figure}[b]
\vspace{-0.2cm}
\includegraphics[width=9.3cm]{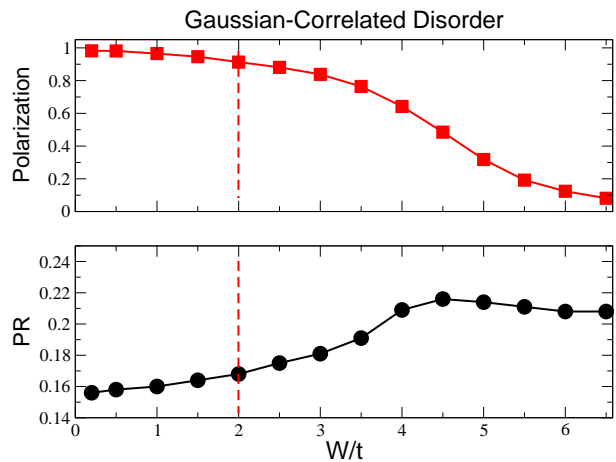}
\vspace{-0.2cm}
\caption{(color online) Values of valley polarization and participation ratio (PR) for the states at Dirac point (at the center of the $n$=0 LL), as a function of Gaussian-correlated disorder amplitude. Dashed line indicates the $W/t$ for which the broadening of LLs is such that the $n$=0 LL starts to overlap with neighbors bands.}
\label{Fig4}
\end{figure}

In Figures 5 and 6 we show in more detail the behavior of the wavefunctions at the center of the $n$=0 LL with increasing disorder, for white-noise and for correlated disorder, respectively. The graphs show the sublattice polarization (upper panels) and the participation ratio (lower panels) for the delocalized state at $E$=0  (we take in fact mean values of the group of states closest to $E$=0 to minimize the small fluctuations still present even when averages over hundreds of disorder realizations are undertaken).  The anomalous increase of PR with disorder we pointed out in Fig.1 is more clearly appreciated in Figs.5 and 6. The important point here is that the evolution of the PR can be simultaneously followed with the valley polarization of these wavefunctions. In this way, a correspondence between the degree of localization and valley (sublattice) polarization can be fully appreciated in these figures, as well as the qualitative difference between both disorder models.  
For both models, the anomalous increase in the PR is accompanied by an equivalent rate of decrease in the polarization.  Vertical dashed lines indicate the disorder amplitude $W/t$ for which the $n$=0 Landau band begin to overlap with the $n$=$\pm$1 bands (the band tails start to touch each other) within each model. For white-noise, Fig.5, the polarization drops faster and is almost  suppressed at $W/t=1$. Just before this threshold in the density of states landscape, the PR reaches its maximum and for higher degree of disorder the localization recovers the usual behavior of the other Landau levels. On the other hand, for Gaussian correlated disorder, Fig. 6, the overall picture seems to be more involved. The polarization is almost unaffected at $W/t=2$, the Landau band overlap threshold for this disorder model. It should be noticed that the polarization is still appreciable for disorder strengths at which any modulation in the density of states are completely washed out. Thi
 s is the case for $W/t>3.5$ (see Fig.1b), but the polarization at $W/t=4.0$ is not negligible (around 65 $\%$). The observation that the PR
continues to increase up to these high disorder values seems to be related to the fact that the
valley polarization is still appreciable for this disorder.

%%%%% Conclusions %%%%%%%%%%%%%%%%%%%%%%%

In conclusion, these results represent a strong evidence of connection between the anomalous localization properties observed for the $n$=0 LL and the way that disorder frustrates the valley polarization in this Landau level. As disorder increases in the graphene system, sublattice-polarized wavefunctions from the $n$=0 LL increasingly spread to the other sublattice, allowing more sites to participate in the wavefunction. This increases, by definition, the participation ratio of the state. The expected and well known localization with increasing disorder only comes back to this scenario when disorder has significantly destroyed the sublattice polarization. Another striking found is that the polarization suppression, as a function of energy within the lowest LL, is model dependent, with the intriguing ``{\it U-shape}" for the white-noise model. On the other hand, the increasing LL mixing also induces a partial polarization in the higher levels.
These observations may establish a key observable for unveiling the striking differences between the
properties of the Dirac-like in comparison to Schroedinger-like
QHE.

We are grateful to A.H. Castro Neto for stimulating discussions and L. Brey for calling the attention to
the problem. This work is supported by FAPESP and CNPq.

%%%%% References  %%%%%%%%%%%%%%%%%%%%%%%

\end{document}